\documentclass[structabstract]{aa}  
\usepackage{graphicx}
\usepackage{txfonts}
\usepackage{longtable}
\usepackage{natbib}
\bibpunct{(}{)}{;}{a}{}{,}
\begin{document}
   \title{VLBA monitoring of Mrk 421 at 15\,GHz and 24\,GHz during 2011}


   \author{R. Lico\inst{1,2}\fnmsep\thanks{Email: rocco.lico@studio.unibo.it}, M. Giroletti\inst{1}, M. Orienti\inst{1,2}, G. Giovannini\inst{1,2}, W. Cotton\inst{3}, P. G. Edwards\inst{4}, L. Fuhrmann\inst{5}, \\ T.P. Krichbaum\inst{5}, K.V. Sokolovsky\inst{6,7}, Y.Y. Kovalev\inst{6,5}, S. Jorstad\inst{8,9}, A. Marscher\inst{8}, M. Kino\inst{10}, D. Paneque\inst{11}, \\ M.A. Perez-Torres\inst{12} \and G. Piner\inst{13}.       
           }

   \institute{INAF Istituto di Radioastronomia, via Gobetti 101, 40129 Bologna, Italy
\and
Dipartimento di Astronomia, Universit\`a di Bologna, via Ranzani 1, 40127 Bologna, Italy
\and             
National Radio Astronomy Observatory, Charlottesville, 520 Edgemont Road, VA 22903-2475, USA
\and
CSIRO Australia Telescope National Facility, Locked Bag 194, Narrabri NSW 2390, Australia
\and
Max-Planck-Institut f\"ur Radioastronomie, Auf dem H\"ugel 69, D-53121 Bonn, Germany
\and
Astro Space Center of Lebedev Physical Institute, Profsoyuznaya 84/32, 117997 Moscow, Russia
\and
Sternberg Astronomical Institute, Moscow State University, Universitetskij prosp. 13, 119992 Moscow, Russia
\and
Institute for Astrophysical Research, Boston University, 725 Commonwealth Avenue, Boston, MA 02215, USA
\and
Astronomical Institute, St. Petersburg State University, Universitetskij Pr. 28, 198504 St. Petersburg, Russia
\and
National Astronomical Observatory of Japan, Osawa 2-21-1, Mitaka, Tokyo 181-8588
\and
Max-Planck-Institut f\"ur Physik, F\"ohringer Ring 6, D-80805 M\"unchen, Germany
\and
Instituto de Astrof\'{\i}sica de Andalucia, IAA-CSIC, Apdo. 3004, 18080 Granada, Spain
\and
Department of Physics and Astronomy, Whittier College, 13406 E. Philadelphia Street, Whittier, CA 90608, USA
            }


 
  \abstract
   {High-resolution radio observations are ideal for constraining the value of physical parameters in the inner regions of active-galactic-nucleus jets and complement results on multiwavelength (MWL) observations. This study is part of a wider multifrequency campaign targeting the nearby TeV blazar Markarian 421 (z=0.031), with observations in the sub-mm (SMA), optical/IR (GASP), UV/X-ray (Swift, RXTE, MAXI), and $\gamma$ rays (Fermi-LAT, MAGIC, VERITAS).}
   {We investigate the jet's morphology and any proper motions, and the time evolution of physical parameters such as flux densities and spectral index. The aim of our wider multifrequency campaign is to try to shed light on questions 
such as the nature of the radiating particles, the connection between the radio and
$\gamma$-ray emission, the location of the emitting regions and the origin of the flux variability.}
   {We consider data obtained with the Very Long Baseline Array (VLBA) over 
   twelve epochs (one observation per month from January to December 2011) at 15\,GHz and 24\,GHz. We investigate the inner jet structure on parsec scales through the study of model-fit components for each epoch.}
   {The structure of Mrk 421 is dominated by a compact ($\sim$0.13 mas) and bright component, with a one-sided jet detected out to $\sim$10 mas. We identify 5-6 components in the jet that are consistent with being stationary during the 12-month period studied here. Measurements of the spectral index agree with those of other works: they are fairly flat in the core region and steepen along the jet length. Significant flux-density variations are detected for the core component.}
   {From our results, we draw an overall scenario in which we estimate a viewing angle $2^\circ < \theta < 5^\circ$ and a different jet velocity for the radio and the high-energy emission regions, such that the respective Doppler factors are $\delta_r \sim 3$  and $\delta_{\rm h.e.} \sim 14$.}

   \keywords{Galaxies: active -- 
                BL~Lacertae objects: Mrk 421  --
                Galaxies: jets
                 }
\authorrunning{R. Lico et al.}
\titlerunning{VLBA monitoring of Mrk 421 at 15\,GHz and 24\,GHz during 2011}

   \maketitle
%

\section{Introduction}

\begin{table*}
\caption{Details of the observations.\label{observations}}
\begin{center}
\tiny
\begin{tabular}{ccccccccc}
\hline
Observation & MJD &  \multicolumn{2}{c}{Map peak} & \multicolumn{2}{c}{Beam} & \multicolumn{2}{c}{1$\sigma$ rms} 
& Notes  \\
date  & & \multicolumn{2}{c}{(mJy/beam)} & \multicolumn{2}{c}{(mas $\times$ mas, $^\circ$)} &
\multicolumn{2}{c}{(mJy/beam)}  &  \\
     & & 15\,GHz & 24\,GHz & 15\,GHz & 24\,GHz& 15\,GHz & 24\,GHz&   \\
\hline
\hline
2011/01/14 & 55575 & 348 & 319 &\ \ 1.05 $\times$ 0.65, 15.3  \ \ &\ \ 0.79 $\times$ 0.47, 8.84 \ \ & 0.19 & 0.18 & No MK, no NL    \\
2011/02/25 & 55617 & 391 & 338 & 1.16 $\times$ 0.74, 14.4 & 0.64 $\times$ 0.39, $-$6.52 & 0.35 & 0.18 & NL snowing \\
2011/03/29 & 55649 & 386 & 359 & 1.06 $\times$ 0.66, $-$5.37 & 0.65 $\times$ 0.39, $-$4.26 & 0.17 & 0.24 & No HK \\
2011/04/25 & 55675 & 367 & 308 & 0.92 $\times$ 0.50, $-$3.79 & 0.61 $\times$ 0.34, $-$3.88 & 0.28 & 0.33 & -  \\
2011/05/31 & 55712 & 355 & 297 & 0.93 $\times$ 0.51, $-$5.56 & 0.64 $\times$ 0.35, $-$8.41 & 0.25 & 0.29 & - \\
2011/06/29 & 55741 & 262 & 208 & 0.89 $\times$ 0.50, $-$7.17 & 0.56 $\times$ 0.32, $-$12.3 & 0.17 & 0.37 & No LA   \\
2011/07/28 & 55770 & 220 & 197 & 0.91 $\times$ 0.56, $-$0.89 & 0.60 $\times$ 0.37, $-$1.14 & 0.20 & 0.30 & -  \\
2011/08/29 & 55802 & 275 & 200 & 0.97 $\times$ 0.55, 0.26 & 0.63 $\times$ 0.35, $-$2.70 & 0.18 & 0.27 & No HK  \\ 
2011/09/28 & 55832 & 264 & 238 & 1.06 $\times$ 0.67, 16.2 & 0.72 $\times$ 0.47, 18.4 & 0.26 & 0.24 & No MK  \\ 
2011/10/29 & 55863 & 261 & 167 & 1.06 $\times$ 0.69, 0.09 & 0.73 $\times$ 0.44, $-$3.73 & 0.26 & 0.14 & HK snowing  \\ 
2011/11/28 & 55893 & 283 & 201 & 1.02 $\times$ 0.59, 18.0 & 0.70 $\times$ 0.42, 14.3 & 0.28 & 0.17 & No PT, FD, MK  \\ 
2011/12/23 & 55918 & 295 & 287 & 0.89 $\times$ 0.49, 1.82 & 0.61 $\times$ 0.35, $-$5.74 & 0.15 & 0.21 & No HK  \\ 
\hline
\end{tabular}
\end{center}
\end{table*}

\begin{figure*}
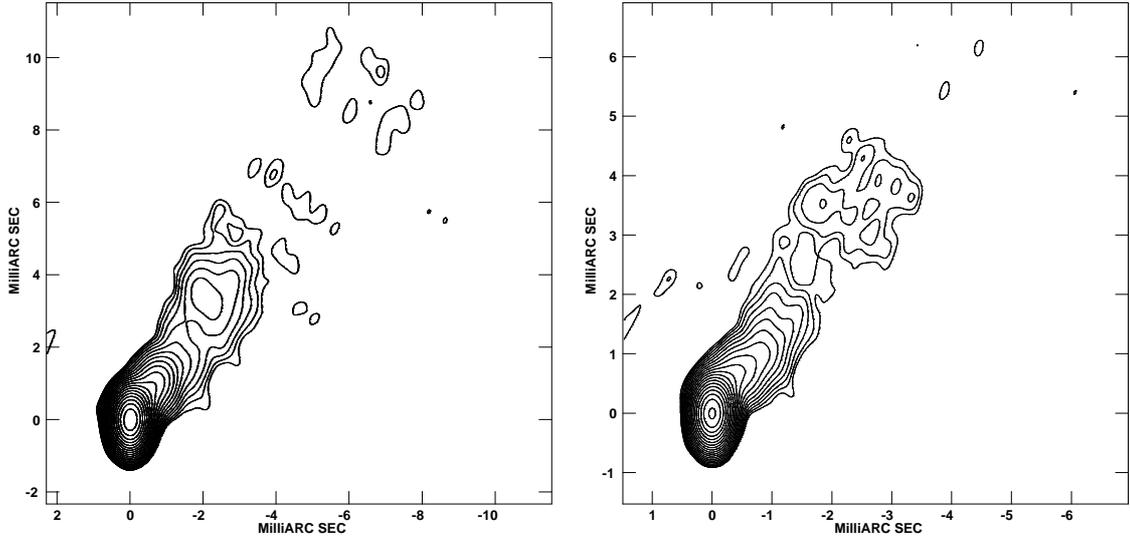

 \centering
 \includegraphics[width=7.5cm]{15ghz_stacked.eps} 
 \includegraphics[width=7.5cm]{24ghz_stacked.eps}
 \caption{Images of Mrk 421 at 15\,GHz (left panel) and 24\,GHz (right panel). These two images were obtained by stacking all of the images of the twelve epochs, at the respective frequency. The restoring beam for the 15\,GHz image is 0.9 mas$\times$0.55 mas and the peak flux density is 307.2 mJy/beam. For the 24\,GHz image, the restoring beam is 0.6 mas$\times$0.35 mas and the peak is 251.9 mJy/beam. The first contour is 0.35 mJy/beam, which corresponds to three times the off-source noise level. Contour levels are drawn at ($-$1, 1, 1.4, 2, 2.8, 4...) in steps of $\sqrt{2}$.}
\label{maps}
 \end{figure*}

Markarian 421 (R.A.=$11^h$\ $04^m$\ $27.313943^s$, Dec.=$+38^\circ$\ 12\arcmin\ 31.79906\arcsec, J2000) is one of the nearest ($z=0.031$) and brightest BL~Lac objects in the sky. It was the first extragalactic source detected at 
TeV energies by the Cherenkov telescope at Whipple Observatory \citep{Punch1992}. The spectral energy 
distribution (SED) of this object, dominated by non-thermal emission, has two
smooth broad components: one at lower energies, from radio band to the soft X-ray domain, and another at higher
energies peaking at $\gamma$-ray energies \citep{Abdo2011}. 
The low-frequency peak is certainly due to synchrotron emission from relativistic electrons in the jet interacting 
with the magnetic field, whereas the high-frequency peak is probably due to the inverse Compton scattering of the same population of 
relativistic electrons with synchrotron low-energy photons \citep[synchrotron-self-compton model, SSC, see][]{Abdo2011, Tavecchio2001}. 
In this framework, multiwavelength (MWL) coordinated campaigns are a fundamental tool for understanding the physical properties of the source, e.g. by studying variability, which is present at all frequencies, but particularly TeV energies where \citet{Gaidos1996} measured a doubling time of $\sim15$ minutes. The accurate MWL study and SED modeling performed by \citet{Abdo2011} revealed some interesting results, such as the size of the emitting region $R$ and the magnetic field $B$, which in the context of the leptonic scenario they constrained to be $R \lesssim 10^4 R_g$ and $B \sim$ 0.05 G. 
However, the details of the physical processes responsible for the observed emission are still poorly constrained.
Because of the considerable variability and the broadband spectrum, multiwavelength  long-term observations are required for a good comprehension of the emission mechanisms.
 
This study is part of a new multi-epoch and multi-instrument campaign, which also involves observations in the sub-mm (SMA), optical/IR (GASP), UV/X-ray (Swift, RXTE, MAXI), and $\gamma$ rays (Fermi-LAT, MAGIC, VERITAS), as well as at the cm wavelengths with low resolution observations (e.g. F-GAMMA, Medicina). 
The aim of this observational effort is to shed light on fundamental questions such 
as the nature of the radiating particles, the connection between the radio and $\gamma$-ray emission, 
the location of the emitting regions, and the origin of the flux variability.
Very long baseline interferometry (VLBI) plays an important role in addressing these scientific questions because it is the only technique that can resolve (at least partially) the inner structure of the jet.
Therefore, cross-correlation studies of Very Long Baseline Array (VLBA) data with data from other energy ranges (in particular $\gamma$ rays) 
can provide us with important information about the structure of the jet and the location of the blazar emission.

At radio frequencies, Mrk 421 clearly shows a one-sided jet structure aligned at a small angle with respect 
to the line of sight \citep{Giroletti2006}.
In this work, we present new VLBA observations to study in detail the inner jet
structure on parsec scales. We are able to investigate the evolution of shocks that arise in the jet, 
by means of the model-fitting technique. In earlier works \citep{Piner1999, Piner2004}, the jet
components show only subluminal apparent motion, which seems to be a common
characteristic of TeV blazars. Thanks to accurate measurements of changes on parsec scales, by the VLBA, we can find
valid constraints on the geometry and kinematics of the jet. 

This paper is structured as follows: in Section 2 we introduce the dataset, in Section 3 we report the results of this work (model fits, flux density variations, apparent speeds, jet sidedness, spectral index), and in Section 4 we discuss results giving our own interpretation in the astrophysical context. We used the following conventions for cosmological parameters:  $H_0=70$ km sec$^{-1}$ Mpc$^{-1}$, $\Omega_M=0.25$ and $\Omega_\Lambda=0.75$, in a flat Universe. We defined the spectral index $\alpha$ such that $S_{\nu}\propto\nu^\alpha$.

\begin{figure*}
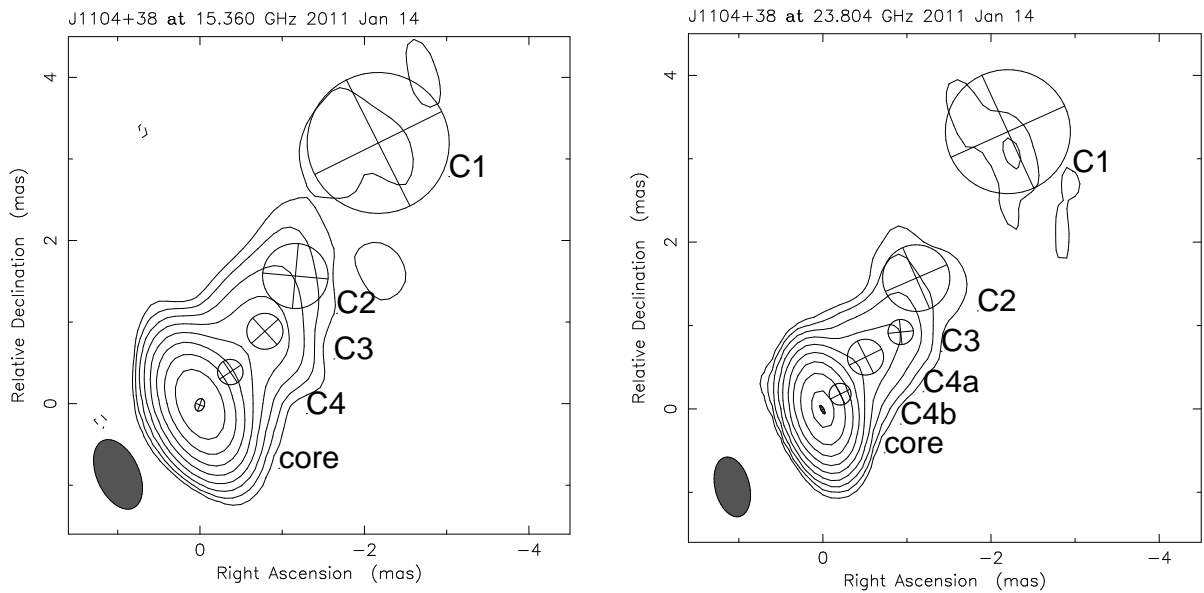

\centering
\includegraphics[angle=-90, clip, width=0.9\columnwidth]{gen15.ps}  
\includegraphics[angle=-90, clip, width=0.9\columnwidth]{gen24.ps} \\
\caption{Images of Mrk 421 with model fit components for the first epoch at 15\,GHz (left panel) and at 24\,GHz (right panel). Levels are drawn at $(-1, 1, 2, 4...) \times$ the lowest contour, that is at 1.0 mJy/beam for both images, in steps of 2. The restoring beam is shown in the bottom left corner; its size is given in Table~\ref{observations}.} 
\label{components}
\end{figure*}

\section{Observations}
We observed Mrk 421 throughout 2011 with the VLBA. 
The source was observed once per month, for a total of 12 epochs, at three 
frequencies: 15, 24, and 43 GHz. In this paper, we present the complete analysis of the whole 15 and 24\,GHz datasets.
We also observed, at regular intervals, three other sources 
(J0854+2006, J1310+3220, and J0927+3902) used as fringe finders and calibrators for the band pass, the instrumental (feed) polarization, and the electric-vector position angle.
At each epoch, Mrk 421 was observed for nearly 40 minutes at each frequency, spread into several scans of about 3 minutes each, interspersed  with calibrator sources in order to improve the (u,v)-coverage. Calibrators were observed for about 10 minutes each, generally spread into three scans of 3 minutes. 

For calibration and fringe-fitting, we used the AIPS software package \citep{Greisen2003}, and for
image production the standard self-calibration procedures included in the DIFMAP software
package \citep{Shepherd1997}, which uses the CLEAN algorithm proposed by \citet{Hogbom1974}. 
In some epochs, one or more antennas did not work properly because of technical
problems; for a complete report (see Table~\ref{observations}).

\section{Results}
\subsection{Images}
We show in Fig.~\ref{maps} two images of Mrk 421 at 15 and 24\,GHz, which were produced by stacking all the images respectively at 15 and 24\,GHz created with DIFMAP. The alignment of the images was checked by comparing the pixel position of the peak. In both images, we set the lowest contour equal to about three times the off-source residual rms noise level.

All the 12 images at each frequency show a similar structure, consisting of a well-defined and well-collimated one-sided jet structure emerging from a compact nuclear region (core-dominated source). This is the typical structure of a BL~Lac object
\citep{Giroletti2004a}. 
The jet extends for roughly 4.5 mas (2.67 pc)\footnote{1 mas corresponds to 0.59 pc.}, 
with a position angle (PA) of $\sim-35^\circ$ (measured from north through east). This morphology agrees with the results of other studies of similar angular resolution \citep{Marscher1999}.

Since the morphology was very stable from epoch to epoch, the stacking did not smear any details of the structure. A sample couple of 15 and 24\,GHz images is shown in Fig.~\ref{components}.
\subsection{Model fits}
For each epoch, we used the model-fitting routine in DIFMAP to fit the visibility data of the source in the $(u,v)$ plane with either elliptical or circular Gaussian components. In this way, we were able to investigate in detail the inner jet structure and its evolution. 
For all epochs at 15\,GHz, a good fit was obtained with five Gaussian components, while at 24\,GHz we needed six components. At both frequencies, we identified the core with the brightest, innermost, and most compact feature.
We label the other components C1, C2, C3, and C4, starting from the outermost (C1) to the innermost (C4). 
The higher angular resolution achieved at 24\,GHz resolves the second innermost 15\,GHz component (C4 located at $\sim$0.45 mas from the core) into two features (C4b at $\sim$0.3 mas and C4a $\sim$0.7 mas from the core) (see Fig.~\ref{components}).

\begin{figure}
\centering
\includegraphics[clip,width=\columnwidth]{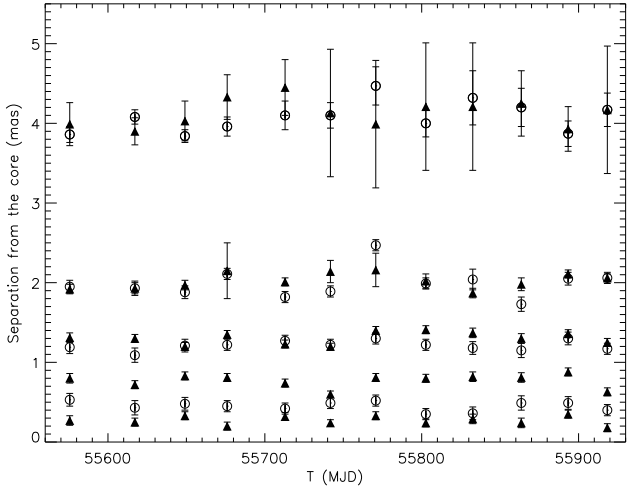} 
\caption{Results of model-fit analysis. Circles and triangles refer, respectively, to positions of the Gaussian components at 15 and 24\,GHz.} 
\label{modelfit}
\end{figure}

Thanks to the extremely fine time-sampling, we were able to make an attempt to identify the same component in each epoch.
Overall, the components extend out to a region of about 5 mas. In this way, with a limited number of components, it was possible to analyze the proper motions and flux density levels at various times. 
From Fig.~\ref{modelfit}, we can clearly see that the data occupy well-defined regions in the radius vs. time plot, and that this behavior helps us to identify the individual components across epochs. 

All details of the model fit analysis are shown in Table~\ref{gaussian}.
We calculated the uncertainties in the position (error bars in Fig.~\ref{modelfit}) using the ratio of the size of each component to the signal-to-noise ratio (S/N). In the case of very bright, compact components, the nominal error value is too small and replaced by a conservative value equal to 10\% of the beam size \citep{Orienti2011}. On the other hand, when the calculated error is very large (i.e. comparable to the component radius, as in the case of a very extended component with a very low flux density), we replace it with a threshold value equal to the maximum value of the calculated errors for the same component at different epochs.

By comparing with \citet{Piner2010} and \citet{Piner2005}, we can argue that the components C2, C3, C4a, and C4b of the present analysis quite likely correspond respectively to the components C5, C6, C7, and C8 of the aforementioned works. We also suggest that our component C1 corresponds to their component C4a. We will consider a more accurate identification of our components with those of previous analyses in a forthcoming paper, where the 43\,GHz data analysis will also be presented.

\addtocounter{table}{1}

\subsection{Flux density variability} \label{flux}
Using the results of the model-fit technique, we analyzed the temporal evolution of
the flux density for each component of the source. The brightest component represents the core; at 15\,GHz, it has a mean value of around 350 mJy, which decreases along the jet until values of about 10 mJy. By comparing the flux density of each component at the various epochs, it emerges that there are no significant variations in the flux densities of the C1-C4 components. The flux density of each component remains roughly constant
at various times within the uncertainties calculated; in any case, there is no indication of flaring activity. 
Any small variations may be artifacts brought about by our fitting procedures: for instance, the flux density of the inner components may be underestimated in some cases, because part of it was incorporated into the core component flux, or the flux density of the most extended features may be underestimated at epochs missing some short baseline data because of telescope failures.
This is a remarkable result because it also allows us to identify components based on their flux density values and confirms our choice of components based on their positions.

\begin{figure}
\centering
\includegraphics[width=\columnwidth]{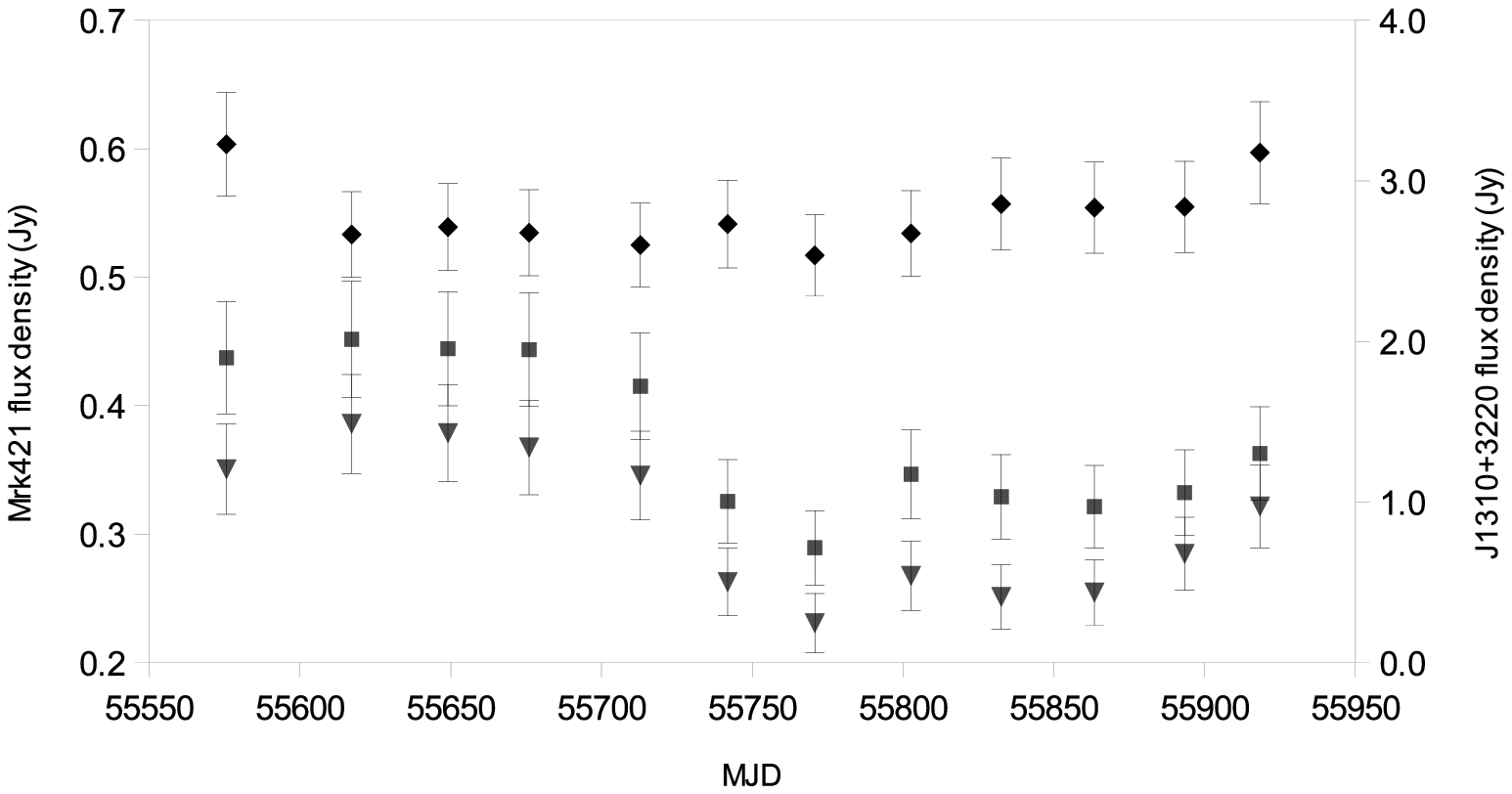} \\ 
\includegraphics[width=\columnwidth]{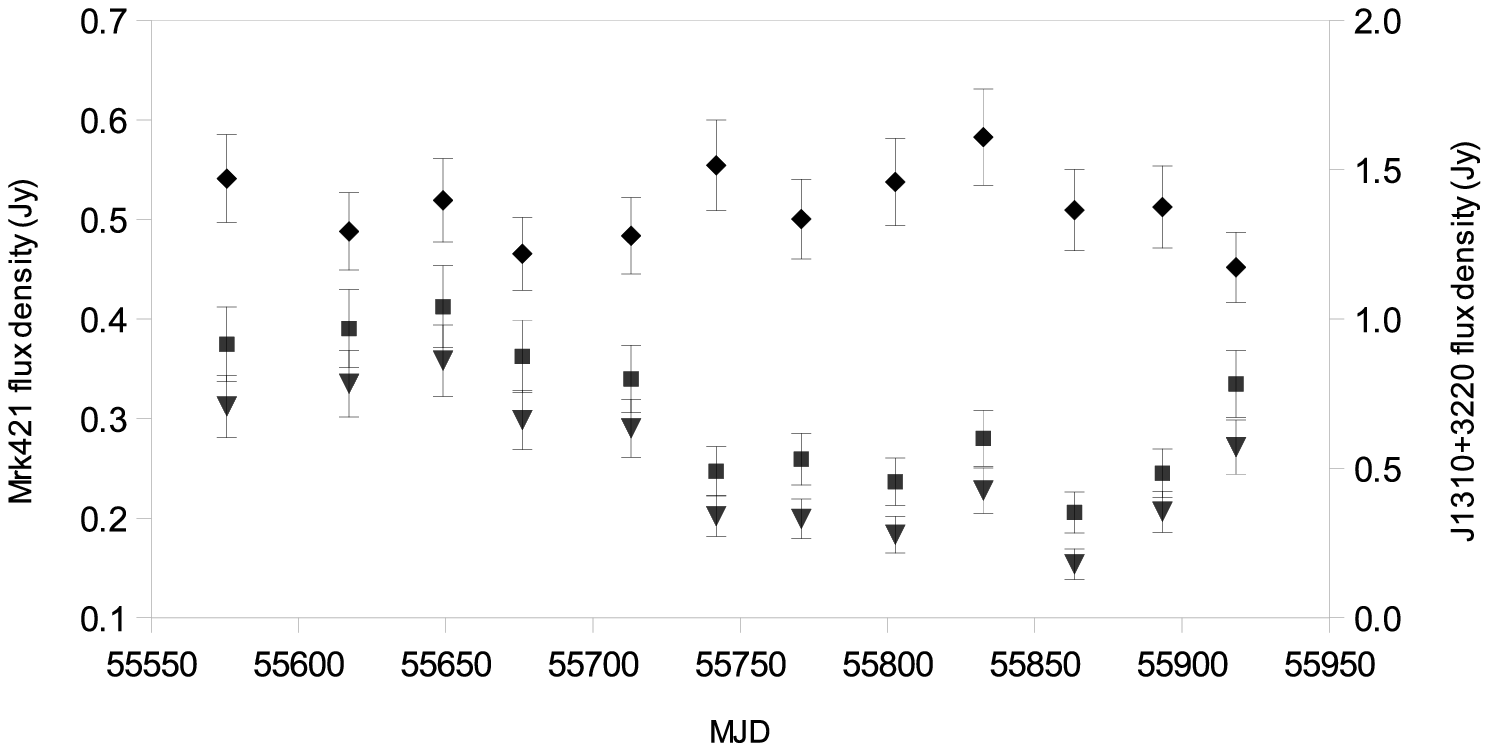}
\caption{Light curves for Mrk421 (squares represents the total flux density and triangles represent the core flux density) and the calibrator J1310+3220 (diamonds represent the total flux density, with scale given on the right hand $y$-axis). The upper and lower panels refer respectively to the 15\,GHz and 24\,GHz data.} 
\label{lightcurves}
\end{figure}

In Fig.~\ref{lightcurves}, we show the light curve for Mrk 421 during 2011 at 15\,GHz (upper panel) and at 24\,GHz (lower panel), considering the total flux density (squares) and the core flux density (triangles). The light curve reveals an interesting feature: in the second part of the year (starting at MJD $\sim$55700), we clearly note a decrease in the total flux density. From the complete flux-density analysis, we found that the core is the component responsible for the decrease, while the extended region does not display any significant variations. To further exclude calibration effects, we performed the same analysis on the three calibrators. In Fig.~\ref{lightcurves}, we present the light curves of the calibrator J1310+3220 (diamonds) at 15\,GHz (upper panel) and 24\,GHz (lower panel). From this comparison, we clearly see that the trend of the light curves for the two sources is very different. We can assert that the flux density decrease observed for the core of Mrk 421 is a real feature. Error bars were calculated by considering a calibration error of about $10\%$ of the flux density and a statistical error equal to three times the map rms noise.

\subsection{Spectral index analysis}
Thanks to our multi-frequency data set, we were able to carry out a detailed analysis of the spectral index distribution on the parsec scale. We conducted a quantitative assessment of the spectral index of the components by comparing the flux density at the two frequencies at each epoch and averaging the results to reduce the statistical fluctuations. For the core, we obtained a value of $\alpha\sim-0.3\pm 0.2$, and for the outermost components (C1, C2, and C3, considered altogether), we obtained $\alpha\sim-1.2\pm 0.5$. The uncertainties were calculated from the theory of the propagation of errors using the formula
\begin{equation}
\Delta \alpha = \frac{1}{\log(24/15)} \sqrt{\left(\frac{\Delta S_{15}}{S_{15}}\right)^2 + \left(\frac{\Delta S_{24}}{S_{24}}\right)^2},
\end{equation}
where $S$ and $\Delta S$ represent the flux density and the respective uncertainty (see the last paragraph of section~\ref{flux}).
These values agree with the results obtained from the spectral index image, which we produced with the following procedure: we first produced new images for each epoch and frequency using the same $(u,v)$ range and the same restoring beam ($0.4$ mas $\times$ $0.7$ mas). We then produced the spectral index image by combining the two images, resulting respectively from the stacking of all images at 15\,GHz and at 24\,GHz, using the AIPS task COMB, clipping the pixels with a S/N $<$ 3 using the input images. By summing images we were able to increase the S/N, and results are more reliable; the stability of the individual features guarantees that we do not lose structural information in the averaging process. The alignment of the images at the two frequencies was checked by comparing the pixel position of the peak.
When a shift was present, we used the LGEOM task in AIPS to align the images.
The resulting image presents the typical flat spectrum in the core region, with a steepening along the jet radius. However, despite the higher S/N obtained with image stacking, a significant spread in the spectral index values is present in the diffuse jet region, hence we do not show the image here.


\subsection{Apparent speeds}
From our model-fit, we infer a small or no displacement for the jet components. To verify and quantify this statement, we determined the speeds of each component by means of linear fits to the separation of the individual features from the core at different epochs. For the 
three outer components (C1, C2, and C3), we used combined data at 15 and 24 GHz, since the positions of each component at the two frequencies are generally consistent within the error bars.
For the two inner components (C4a and C4b), we only used data at 24\,GHz.
Results are shown in Table~\ref{speeds}.

We found low values for the apparent speeds, in agreement with previous studies \citep[e.g.][]{Piner2005}.
The two innermost components (C4a and C4b) are essentially stationary, with an upper limit to their separation velocity of $\sim0.1$c. In addition C2 and C3 are consistent with being stationary,  while the outermost component C1 has a low-significance (1.5$\sigma$) subluminal motion$\sim0.3$c. 
If this trend of increasing velocity at larger radii were real and if the apparent speeds shown in Table~\ref{speeds} represented the bulk apparent speed of the plasma in the jet, we could speculate that some mechanism involving an acceleration acts in the outer region of the jet.

\begin{table}[t]
\begin{center}
\footnotesize
\caption{Apparent speeds from the linear fit analysis.\label{speeds}}
\begin{tabular}{lcc}
\hline
Component & Apparent speed & $\beta_{\rm app}$ \\
& (mas/yr) &  \\
\hline
\hline
C1   &$0.17\pm0.12$ & $0.34\pm0.24$ \\
C2   &$0.08\pm0.10$ & $0.16\pm0.20$ \\
C3  &$0.05\pm0.06$ & $0.10\pm0.11$ \\
C4a  &$-0.01\pm0.09$ & $-0.02\pm0.17$ \\
C4b   &$-0.02\pm0.06$ & $-0.03\pm0.11$  \\
\hline
\end{tabular}
\end{center}
\end{table}

\subsection{Jet/counter-jet ratio}
We estimated the ranges of viewing angles $\theta$ and of $\beta$ from the jet/counter-jet brightness ratio. Assuming that the source has two symmetrical jets of the same intrinsic power, we used the equation
$$\frac{B_J}{B_{cJ}}=R=\left(\frac{1+\beta \cos\theta}{1-\beta \cos\theta}\right)^{2-\alpha},$$
where $B_J$ and $B_{cJ}$ are, respectively, the jet and counter-jet brightnesses and $\alpha$ represents the spectral index defined in the introduction; we adopted the $(2-\alpha)$ exponent, since the jet is smooth and does not contain well-defined compact {\it blobs}.
For the jet brightness, we used $B_J\sim28.4$ mJy/beam, measured at 24\,GHz, in the image resulting from the stacking of all the 12 epoch images, in the jet region located at $\sim1$ mas from the core. For the counter-jet, which is not visible, we used an upper limit provided by the $3\sigma$ rms noise level measured in the image, which resulted in $B_{cJ}=0.11$ mJy/beam; this consequently yields a lower limit to  both $R$ and $\beta\cos\theta$. With a value of $\alpha=-0.4$, in agreement with our spectral index images, we obtained $R>254.8$ and then $\beta\cos\theta>0.82$. Therefore, the minimum allowed jet bulk velocity is $\beta_{\rm min}=0.82$ (corresponding to a bulk Lorentz factor $\gamma>1.74$) and the maximum viewing angle is $\theta_{\rm max} = 35.0^\circ$.
\vspace{-0.25 cm}
\section{Discussion and conclusions}

The Doppler factor defined as

$$\delta=\frac{1}{\gamma(1-\beta \cos\theta)}$$

\noindent is a key element in the study of blazars, since it affects various parameters such as the observed brightness, the SED peak frequency, the variability timescale, and more. Modeling of the SED and study of the variability in different wavebands generally require large values of the blazar Doppler factors; in the case of Mrk421, \citet{Gaidos1996} estimated $\delta>9$, from the observed TeV variability time of about 30 min and \citet{Abdo2011} required a Doppler factor between 20 and 50 to reproduce the broadband SED. In turn, VLBI observations can also constrain $\delta$ by posing limits on $\beta$ and $\theta$, as provided by the various arguments discussed in this work.

When closely spaced repeated observations are available, the study of the proper motion is a useful tool in determining the ranges for $\beta$ and $\theta$. Surprisingly, several works \citep[e.g.][]{Giroletti2004b, Piner2008, Piner2010} have reported subluminal motions, sometimes consistent with the component being stationary, in the jets of all the $\sim$10 TeV blazars for which 
proper motion studies have been performed in the literature. Thanks to the large number of dual-frequency observations, the fine time-sampling, and the high quality of the data provided by the good $(u,v)$-coverage, we performed a robust identification of the Gaussian components and constrained their motion to be consistent with no displacement at all. At the same time, the high sensitivity and in particular the stacked image place significant constraints on the jet/counter-jet ratio.

The first immediate consequence is that we can reject the hypothesis that the small $\beta_{\rm app}$ is solely due to a projection effect, since it would require an unrealistically narrow viewing angle: in the case of component C4b, the upper limit to the observed motion implies a viewing angle $<1.3^\circ$ to reproduce the observed jet/counter-jet ratio (and even smaller to agree with the high energy limits). If the jets' distribution is isotropic on the sky, the real number of misaligned sources (parent population) is incompatible with these very small values of $\theta$. For example, in the Bologna Complete Sample selected by \citet{Giovannini2005} at low frequency (and thus free from Doppler favoritism bias), one would expect fewer than 0.03 sources with $\theta<1.3^\circ$ \citep[see also][]{Tavecchio2005,Henri2006}. 

On the other hand, since larger values of the viewing angle capable of reproducing the observed lack of proper motion are incompatible with the jet/counter-jet ratio, we conclude that the pattern velocity cannot be representative of the bulk flow velocity. The Gaussian components obtained in our model fit provide a good description of the visibility data but do not represent well-defined, high-contrast jet features \citep[see][]{Lyutikov2010}. In our interpretation, the low apparent speeds found imply that the proper motion of Mrk421 does not provide any information about the jet bulk velocity; even on the basis of the sole jet brightness ratio, untenable viewing angles would be necessary to match the pattern and bulk velocities. 

What then are the real values of the viewing angle and the jet bulk velocity in Mrk421? To reproduce the observed jet asymmetry, we needed to consider a range of velocities $0.82<\beta<1$ and angles $0^\circ<\theta<35.0^\circ$. 
We could exclude the upper range of the $\theta$ values, since this would not reproduce the high Doppler factors required by high energy observations \citep{Gaidos1996, Abdo2011}; in particular, we were unable to achieve $\delta>20, 10, 5, 3$ when $\theta>3.0^\circ, 5.7^\circ, 11.5^\circ, 19.4^\circ$, respectively.

Smaller angles thus seem to be favored since they were the only ones consistent with high values of $\delta$; however, such small angles would still represent a challenge to the observed radio properties. We were able to estimate the intrinsic power of the radio core $P_c^{\rm intr}$, by debeaming the observed monochromatic luminosity  of the core $P_c^{\rm obs}$ with the equation

$$P_c^{\rm obs}=P_c^{\rm intr} \times \delta_c^{2-\alpha}.$$
With a value of $P_c^{\rm obs}\sim6.8\times10^{23}$ W\,Hz$^{-1}$ at 15\,GHz,  $\alpha=-0.3$, and $\delta=20$, we obtained for the intrinsic power of the core $P_c^{\rm intr}\sim5.8\times10^{20}$ W\,Hz$^{-1}$; this value was at the very low end of the typical range of intrinsic power found for different samples of radio galaxies \cite[e.g.][]{Liuzzo2011}, suggesting that lower values of $\delta$ provide a more typical core power. 

Moreover, the 12 monthly observations have not revealed any dramatic flux-density variability in the core of the source, which further points to a lower Doppler factor for the radio jet.
We estimated the variability brightness temperature of the core ($T_{\rm B,var}$) with the formula proposed by \citet{Hovatta2009}
$$ T_{\rm B,var}= 1.548 \times 10^{-32} \frac{\Delta S_{\rm max}d_L^2}{\nu^2\tau^2(1+z)},$$
where $\nu$ is the observed frequency in GHz, $z$ is the redshift, $d_L$ is
the luminosity distance in meters, $\Delta S_{\rm max}$ is the difference between the maximum value for the core flux density and the minimum value, and $\tau$ is the variability time. With the values provided by our observations and $\tau$ $\sim$ 90 days, we obtaind a value of $T_{\rm B,var} \sim 2.1\times 10^{10}$~K, which does not require any significant beaming.

We similarly calculated $T_{\rm B}$ for the most compact component following the standard formula \citep{Piner1999,Tingay1998}
$$T_{\rm B}=1.22 \times 10^{12} \frac{S(1+z)}{ab\nu^2},$$
where $S$ is the flux density of the component measured in Jy, $a$ and $b$ are the full widths at half maximum of the major and minor axes respectively of the component measured in mas, $z$ is the redshift, and $\nu$ is the observation frequency in GHz. The resulting $T_{\rm B}$'s are on the order of a few $\times 10^{11}$ K, only slightly exceeding the limit derived by \citet{Readhead1994} from equipartition arguments.

Taken together, the lack of superluminal features, the low core dominance, and the weak variability suggest a scenario in which no strong beaming is required in the radio jet. This is not uncommon in TeV blazars \citep{Piner2004, Piner2008}, but unprecedentedly firm observational support for it has been provided by our intensive campaign. Low values of the Doppler factor, e.g. $\delta \sim 3$, can reproduce the observational radio properties, including the jet brightness asymmetry. 

We conclude that the Doppler factor must be different in the radio band than the $\gamma$-ray band. Since we do not expect that the viewing angle changes significantly, this leads us to the necessity of a velocity structure in the jet, as previously discussed by e.g.\ \citet{Chiaberge2000},  \citet{Georganopoulos2003}, and \citet{Ghisellini2005}. Our images do not provide strong evidence in favour of either a radial or transverse velocity structure, although previous works have revealed a limb brightening in Mrk421, on both a milliarcsecond scale at 43 GHz \citep{Piner2010} and at $d>10$ mas at 5 GHz \citep{Giroletti2006}. This would favor the presence of a transverse velocity structure across the jet axis. This structure consists of two components: a fast inner \textit{spine} and a slower outer \textit{layer}. Different Doppler factors were obtained depending on whether we measured the speed of the spine or the layer.  

A viable scenario for Mrk421 is that the viewing angle is between $2^\circ$ and $5^\circ$, which is consistent with the statistical counts of low-power radio sources and the possibility of reaching the high Doppler factors required by SED modeling and high-energy variability. The jet velocity is structured, with a typical Lorentz factor of $\gamma\sim 1.8$ in the radio region (yielding $\delta \sim 3$), and $\gamma \sim \delta \sim 20$ in the high-energy emission region. For example, assuming  $\theta=4^\circ$, $\beta_{\rm h.e.}=0.998$, and $\beta_{\rm r}=0.82$, we obtained a value of $\delta_{\rm h.e.}=14.3$ and $\delta_{\rm r}=3.2$ and successfully reproduced all the observational properties of the source.

In summary, the detailed analysis presented in this paper has largely confirmed with improved quality expectations based on the knowledge so far achieved for TeV blazars. We have also estimated with a good level of significance some important and fundamental parameters ($\delta$, $\theta$, $\beta$, $\alpha$) that characterize the physical processes in blazars. 
However, there is still much to be understood and we expect to obtain other
significant results from the analysis extended to other wavelengths, particularly in the $\gamma$-ray domain. Additional works on the dataset presented in this paper are planned, and will deal with, e.g., the 43\,GHz images and the polarization properties. Moreover, we intend to combine our dataset  with those of other works (e.g.\ \citealt{Piner2005} or the MOJAVE survey, \citealt{Lister2009}) to increase the temporal coverage of the observations and obtain even tighter constraints over a longer time frame.
\vspace{-0.2 cm}
\begin{acknowledgements}
This work is based on observations obtained through the BG207 VLBA project, which makes use of the Swinburne University of Technology software correlator, developed as part of the Australian Major National Research Facilities Programme and operated under licence \citep{Deller2011}. The
National Radio Astronomy Observatory is a facility of the National Science Foundation operated
under cooperative agreement by Associated Universities, Inc. For this paper we made use of the NASA/IPAC Extragalactic Database NED which is operated by the JPL, Californian Institute of Technology, under contract with the National Aeronautics and Space Administration. We acknowledge financial contribution from grant PRIN-INAF-2011. This research is partially supported by KAKENHI (24540240). KVS and YYK are partly supported by the Russian Foundation for Basic Research (project 11-02-00368), and the basic research program ``Active processes in galactic and extragalactic objects'' of the Physical
Sciences Division of the Russian Academy of Sciences. YYK is also supported by the Dynasty Foundation. The research at Boston University was supported in part by NASA through Fermi grants NNX08AV65G, NNX08AV61G, NNX09AT99G, NNX09AU10G, and NNX11AQ03G, and by US National Science Foundation grant AST-0907893. We thank Dr. Claire Halliday for the language editing work which improved the text of the present manuscript.
\end{acknowledgements}


\longtab{2}{
\begin{tiny}
\begin{longtable}{lcccccccccc}
\caption{\label{gaussian} Gaussian models.}\\
\hline\hline
Epoch\tablefootmark{a} & Frequency & Component ID & S\tablefootmark{b} & $\sigma_S$\tablefootmark{c} & r\tablefootmark{d} & $\sigma_r$\tablefootmark{e} & PA\tablefootmark{d} & a\tablefootmark{f} & & $\Phi$\tablefootmark{g} \\
 & (GHz) & & (mJy) & (mJy) & (mas) & (mas) & (deg) & (mas) & b/a & (deg) \\
\hline
\endfirsthead
\caption{Continued.}\\
\hline\hline
Epoch\tablefootmark{a} & Frequency & Component ID & S\tablefootmark{b} & $\sigma_S$\tablefootmark{c} & r\tablefootmark{d} & $\sigma_r$\tablefootmark{e} & PA\tablefootmark{d} & a\tablefootmark{f} & & $\Phi$\tablefootmark{g} \\
 & (GHz) & & (mJy) & (mJy) & (mas) & (mas) & (deg) & (mas) & b/a & (deg) \\
\hline
\endhead
\hline
\endfoot
&&&&&&&&&&\\
14/01/11 & 15 & Core & 350 & 35 & ... & ... & ... & 0.16 & 0.71 & -20.0 \\
 &  & C4 & 30 & 3.1 & 0.52 & 0.08 & -43.6 & 0.31 & 1.00 & ... \\
 &  & C3 & 22 & 2.3 & 1.17 & 0.08 & -41.7 & 0.44 & 1.00 & ... \\
 &  & C2 & 8.6 & 1.0 & 1.93 & 0.08 & -36.6 & 0.80 & 1.00 & ... \\
 &  & C1 & 14 & 1.5 & 3.85 & 0.10 & -34.1 & 1.73 & 1.00 & ... \\
 & 24 & Core & 312 & 31 & ... & ... & ... & 0.11 & 0.32 & 27.3 \\
 &  & C4b & 23 & 2.4 & 0.26 & 0.06 & -49.3 & 0.26 & 1.00 & ... \\
 &  & C4a & 17 & 1.8 & 0.79 & 0.06 & -39.2 & 0.43 & 1.00 & ... \\
 &  & C3 & 8.1 & 1.0 & 1.3 & 0.06 & -45.1 & 0.30 & 1.00 & ... \\
 &  & C2 & 7.1 & 0.9 & 1.91 & 0.06 & -35.3 & 0.80 & 1.00 & ... \\
 &  & C1 & 5.9 & 0.8 & 3.98 & 0.27 & -33.5 & 1.49 & 1.00 & ... \\
25/02/11 & 15 & Core & 386 & 39 & ... & ... & ... & 0.10 & 0.65 & 43.7 \\
 &  & C4 & 19 & 2.2 & 0.42 & 0.09 & -44.8 & 0.19 & 1.00 & ... \\
 &  & C3 & 19 & 1.3 & 1.08 & 0.09 & -39.8 & 0.44 & 1.00 & ... \\
 &  & C2 & 8.4 & 1.3 & 1.91 & 0.09 & -44.8 & 0.91 & 1.00 & ... \\
 &  & C1 & 14 & 1.7 & 4.06 & 0.09 & -32.9 & 1.29 & 1.00 & ... \\
 & 24 & Core & 335 & 34 & ... & ... & ... & 0.07 & 1.00 & ... \\
 &  & C4b & 21 & 2.1 & 0.24 & 0.05 & -16.0 & 0.32 & 1.00 & ... \\
 &  & C4a & 14 & 0.9 & 0.71 & 0.05 & -35.6 & 0.36 & 1.00 & ... \\
 &  & C3 & 6.6 & 0.9 & 1.29 & 0.05 & -44.2 & 0.28 & 1.00 & ... \\
 &  & C2 & 5.3 & 0.8 & 1.91 & 0.07 & -34.6 & 0.80 & 1.00 & ... \\
 &  & C1 & 6.7 & 0.9 & 3.89 & 0.17 & -33.9 & 1.17 & 1.00 & ... \\
29/03/11 & 15 & Core & 379 & 38 & ... & ... & ... & 0.04 & 1.00 & ... \\
 &  & C4 & 16 & 1.7 & 0.46 & 0.08 & -34.8 & 0.28 & 1.00 & ... \\
 &  & C3 & 21 & 2.1 & 1.2 & 0.08 & -36.0 & 0.46 & 1.00 & ... \\
 &  & C2 & 4.7 & 0.7 & 1.86 & 0.08 & -45.2 & 0.78 & 1.00 & ... \\
 &  & C1 & 15 & 1.6 & 3.82 & 0.08 & -34.3 & 1.34 & 1.00 & ... \\
 & 24 & Core & 358 & 36 & ... & ... & ... & 0.06 & 1.00 & ... \\
 &  & C4b & 15 & 1.6 & 0.32 & 0.05 & -28.9 & 0.21 & 1.00 & ... \\
 &  & C4a & 13 & 1.5 & 0.83 & 0.05 & -32.7 & 0.25 & 1.00 & ... \\
 &  & C3 & 10 & 1.2 & 1.2 & 0.05 & -44.1 & 0.43 & 1.00 & ... \\
 &  & C2 & 5.7 & 0.9 & 1.96 & 0.06 & -37.3 & 0.72 & 1.00 & ... \\
 &  & C1 & 8.5 & 1.1 & 4.03 & 0.25 & -36.2 & 1.31 & 1.00 & ... \\
25/04/11 & 15 & Core & 367 & 37 & ... & ... & ... & 0.11 & 0.85 & -12.4 \\
 &  & C4 & 22 & 2.4 & 0.45 & 0.07 & -33.6 & 0.25 & 1.00 & ... \\
 &  & C3 & 20 & 2.2 & 1.21 & 0.07 & -36.7 & 0.41 & 1.00 & ... \\
 &  & C2 & 6.7 & 1.1 & 2.11 & 0.07 & -46.6 & 0.85 & 1.00 & ... \\
 &  & C1 & 16 & 1.8 & 3.95 & 0.12 & -33.1 & 1.47 & 1.00 & ... \\
 & 24 & Core & 299 & 30 & ... & ... & ... & 0.05 & 1.00 & ... \\
 &  & C4b & 27 & 2.9 & 0.2 & 0.05 & -41.7 & 0.12 & 1.00 & ... \\
 &  & C4a & 9.4 & 1.4 & 0.81 & 0.05 & -32.6 & 0.26 & 1.00 & ... \\
 &  & C3 & 10 & 1.4 & 1.35 & 0.05 & -39.0 & 0.37 & 1.00 & ... \\
 &  & C2 & 5.2 & 1.1 & 2.14 & 0.35 & -38.6 & 1.04 & 1.00 & ... \\
 &  & C1 & 11 & 1.5 & 4.32 & 0.28 & -33.6 & 1.23 & 1.00 & ... \\
31/05/11 & 15 & Core & 346 & 35 & ... & ... & ... & 0.04 & 1.00 & ... \\
 &  & C4 & 24 & 2.5 & 0.4 & 0.07 & -29.4 & 0.25 & 1.00 & ... \\
 &  & C3 & 17 & 1.8 & 1.26 & 0.07 & -33.6 & 0.40 & 1.00 & ... \\
 &  & C2 & 4.7 & 0.9 & 1.81 & 0.07 & -43.0 & 0.53 & 1.00 & ... \\
 &  & C1 & 15 & 1.7 & 4.09 & 0.18 & -34.4 & 1.72 & 1.00 & ... \\
 & 24 & Core & 290 & 29 & ... & ... & ... & 0.04 & 1.00 & ... \\
 &  & C4b & 18 & 2.0 & 0.31 & 0.05 & -18.0 & 0.18 & 1.00 & ... \\
 &  & C4a & 10 & 1.3 & 0.74 & 0.05 & -31.0 & 0.29 & 1.00 & ... \\
 &  & C3 & 7.5 & 1.1 & 1.23 & 0.05 & -42.1 & 0.41 & 1.00 & ... \\
 &  & C2 & 4.5 & 1.0 & 2.01 & 0.05 & -34.2 & 0.53 & 1.00 & ... \\
 &  & C1 & 8.9 & 1.2 & 4.45 & 0.35 & -35.8 & 1.35 & 1.00 & ... \\
29/06/11 & 15 & Core & 263 & 26 & ... & ... & ... & 0.19 & 0.41 & -16.8 \\
 &  & C4 & 26 & 2.7 & 0.47 & 0.07 & -30.8 & 0.26 & 1.00 & ... \\
 &  & C3 & 16 & 1.7 & 1.2 & 0.07 & -37.0 & 0.51 & 1.00 & ... \\
 &  & C2 & 6.1 & 0.8 & 1.86 & 0.07 & -35.9 & 0.61 & 1.00 & ... \\
 &  & C1 & 19 & 1.9 & 4.08 & 0.16 & -36.8 & 2.01 & 1.00 & ... \\
 & 24 & Core & 202 & 20 & ... & ... & ... & 0.08 & 1.00 & ... \\
 &  & C4b & 27 & 2.9 & 0.23 & 0.04 & -26.3 & 0.20 & 1.00 & ... \\
 &  & C4a & 12 & 1.6 & 0.59 & 0.04 & -28.5 & 0.23 & 1.00 & ... \\
 &  & C3 & 13 & 1.7 & 1.19 & 0.04 & -37.0 & 0.51 & 1.00 & ... \\
 &  & C2 & 6.3 & 1.3 & 2.13 & 0.14 & -39.7 & 0.75 & 1.00 & ... \\
 &  & C1 & 15 & 1.9 & 4.12 & 0.80 & -32.9 & 2.56 & 1.00 & ... \\
28/07/11 & 15 & Core & 231 & 23 & ... & ... & ... & 0.23 & 0.30 & -19.4 \\
 &  & C4 & 26 & 2.6 & 0.5 & 0.07 & -30.6 & 0.30 & 1.00 & ... \\
 &  & C3 & 14 & 1.6 & 1.29 & 0.07 & -35.6 & 0.45 & 1.00 & ... \\
 &  & C2 & 5.5 & 0.8 & 2.45 & 0.07 & -38.3 & 0.99 & 1.00 & ... \\
 &  & C1 & 9.3 & 1.1 & 4.45 & 0.24 & -35.5 & 1.82 & 1.00 & ... \\
 & 24 & Core & 199 & 20 & ... & ... & ... & 0.14 & 0.56 & -30.9 \\
 &  & C4b & 28 & 2.9 & 0.29 & 0.05 & -25.3 & 0.25 & 1.00 & ... \\
 &  & C4a & 10 & 1.4 & 0.77 & 0.05 & -34.4 & 0.41 & 1.00 & ... \\
 &  & C3 & 10 & 1.3 & 1.36 & 0.05 & -35.4 & 0.55 & 1.00 & ... \\
 &  & C2 & 2.2 & 0.9 & 2.13 & 0.21 & -35.4 & 0.70 & 1.00 & ... \\
 &  & C1 & 7.9 & 1.2 & 3.95 & 0.80 & -35.5 & 2.36 & 1.00 & ... \\
29/08/11 & 15 & Core & 267 & 27 & ... & ... & ... & 0.15 & 0.64 & -21.5 \\
 &  & C4 & 30 & 3.1 & 0.33 & 0.07 & -33.3 & 0.38 & 1.00 & ... \\
 &  & C3 & 16 & 1.7 & 1.2 & 0.07 & -32.0 & 0.42 & 1.00 & ... \\
 &  & C2 & 4.6 & 0.7 & 1.97 & 0.07 & -43.9 & 0.64 & 1.00 & ... \\
 &  & C1 & 13 & 1.4 & 3.98 & 0.17 & -34.8 & 1.86 & 1.00 & ... \\
 & 24 & Core & 183 & 18 & ... & ... & ... & 0.03 & 1.00 & ... \\
 &  & C4b & 30 & 3.1 & 0.22 & 0.05 & -18.0 & 0.17 & 1.00 & ... \\
 &  & C4a & 10 & 1.3 & 0.78 & 0.05 & -28.9 & 0.37 & 1.00 & ... \\
 &  & C3 & 4.5 & 0.9 & 1.39 & 0.05 & -40.9 & 0.28 & 1.00 & ... \\
 &  & C2 & 3.6 & 0.9 & 2 & 0.09 & -31.5 & 0.66 & 1.00 & ... \\
 &  & C1 & 6.5 & 1.0 & 4.19 & 0.80 & -41.0 & 2.02 & 1.00 & ... \\
28/09/11 & 15 & Core & 251 & 25 & ... & ... & ... & 0.15 & 0.58 & 12.2 \\
 &  & C4 & 34 & 3.5 & 0.33 & 0.08 & -34.3 & 0.39 & 1.00 & ... \\
 &  & C3 & 18 & 1.9 & 1.15 & 0.08 & -32.1 & 0.59 & 1.00 & ... \\
 &  & C2 & 4.2 & 0.9 & 2.02 & 0.13 & -43.6 & 1.13 & 1.00 & ... \\
 &  & C1 & 12 & 1.5 & 4.3 & 0.34 & -32.9 & 2.26 & 1.00 & ... \\
 & 24 & Core & 228 & 23 & ... & ... & ... & 0.09 & 0.42 & 15.2 \\
 &  & C4b & 24 & 2.5 & 0.27 & 0.06 & -31.0 & 0.15 & 1.00 & ... \\
 &  & C4a & 15 & 1.7 & 0.8 & 0.06 & -30.9 & 0.42 & 1.00 & ... \\
 &  & C3 & 3.3 & 0.8 & 1.35 & 0.06 & -44.4 & 0.55 & 1.00 & ... \\
 &  & C2 & 2.7 & 0.8 & 1.85 & 0.06 & -29.0 & 0.43 & 1.00 & ... \\
 &  & C1 & 9.2 & 1.2 & 4.19 & 0.80 & -33.0 & 2.19 & 1.00 & ... \\
29/10/11 & 15 & Core & 255 & 25 & ... & ... & ... & 0.15 & 0.32 & -33.2 \\
 &  & C4 & 24 & 2.6 & 0.47 & 0.09 & -34.3 & 0.28 & 1.00 & ... \\
 &  & C3 & 17 & 1.8 & 1.12 & 0.09 & -31.7 & 0.44 & 1.00 & ... \\
 &  & C2 & 3.0 & 0.8 & 1.71 & 0.09 & -35.4 & 0.44 & 1.00 & ... \\
 &  & C1 & 11 & 1.4 & 4.18 & 0.24 & -33.0 & 1.98 & 1.00 & ... \\
 & 24 & Core & 154 & 15 & ... & ... & ... & 0.04 & 1.00 & ... \\
 &  & C4b & 21 & 2.1 & 0.23 & 0.06 & -23.2 & 0.18 & 1.00 & ... \\
 &  & C4a & 11 & 1.2 & 0.8 & 0.06 & -31.5 & 0.35 & 1.00 & ... \\
 &  & C3 & 3.3 & 0.5 & 1.29 & 0.06 & -32.8 & 0.41 & 1.00 & ... \\
 &  & C2 & 3.2 & 0.5 & 1.97 & 0.08 & -35.9 & 0.83 & 1.00 & ... \\
 &  & C1 & 8.6 & 1.0 & 4.23 & 0.41 & -33.0 & 1.99 & 1.00 & ... \\
28/11/11 & 15 & Core & 285 & 28 & ... & ... & ... & 0.18 & 0.15 & -14.1 \\
 &  & C4 & 19 & 2.1 & 0.48 & 0.08 & -33.4 & 0.36 & 1.00 & ... \\
 &  & C3 & 14 & 1.7 & 1.29 & 0.08 & -32.3 & 0.41 & 1.00 & ... \\
 &  & C2 & 4.8 & 1.0 & 2.04 & 0.08 & -35.5 & 0.87 & 1.00 & ... \\
 &  & C1 & 4.8 & 1.0 & 3.86 & 0.16 & -33.3 & 1.17 & 1.00 & ... \\
 & 24 & Core & 207 & 21 & ... & ... & ... & 0.14 & 0.58 & -22.7 \\
 &  & C4b & 17 & 1.7 & 0.34 & 0.05 & -24.4 & 0.34 & 1.00 & ... \\
 &  & C4a & 6.7 & 0.8 & 0.87 & 0.05 & -33.7 & 0.36 & 1.00 & ... \\
 &  & C3 & 7.3 & 0.9 & 1.35 & 0.05 & -33.7 & 0.46 & 1.00 & ... \\
 &  & C2 & 2.7 & 0.6 & 2.1 & 0.05 & -32.2 & 0.68 & 1.00 & ... \\
 &  & C1 & 6.0 & 0.8 & 3.92 & 0.28 & -32.4 & 1.44 & 1.00 & ... \\
23/12/11 & 15 & Core & 230 & 23 & ... & ... & ... & 0.07 & 1.00 & ... \\
 &  & C4 & 14 & 1.5 & 0.39 & 0.07 & -27.0 & 0.38 & 1.00 & ... \\
 &  & C3 & 12 & 1.3 & 1.17 & 0.07 & -32.7 & 0.50 & 1.00 & ... \\
 &  & C2 & 5.8 & 0.7 & 2.05 & 0.07 & -35.0 & 0.74 & 1.00 & ... \\
 &  & C1 & 6.2 & 0.8 & 4.16 & 0.21 & -31.4 & 1.53 & 1.00 & ... \\
 & 24 & Core & 271 & 27 & ... & ... & ... & 0.04 & 1.00 & ... \\
 &  & C4b & 30 & 3.1 & 0.16 & 0.05 & -19.5 & 0.16 & 1.00 & ... \\
 &  & C4a & 16 & 1.8 & 0.61 & 0.05 & -28.5 & 0.29 & 1.00 & ... \\
 &  & C3 & 12 & 1.3 & 1.23 & 0.05 & -39.4 & 0.55 & 1.00 & ... \\
 &  & C2 & 6.2 & 0.9 & 2.04 & 0.05 & -35.0 & 0.74 & 1.00 & ... \\
 &  & C1 & 2.6 & 0.7 & 4.15 & 0.80 & -31.4 & 1.53 & 1.00 & ... \\ 
\end{longtable}
\tablefoot{
\begin{tiny}
\newline
\tablefoottext{a}{For more details on various epochs see Table 1.}\\
\tablefoottext{b}{Flux density in mJy.}\\
\tablefoottext{c}{Estimated errors for the component flux density.}\\
\tablefoottext{d}{r and PA are the polar coordinates of the component's center with respect to the core. The position angle (PA) is measured from North through East.}\\
\tablefoottext{e}{Estimated errors in the component position.}\\
\tablefoottext{f}{a and b are the FWHM of the major and minor axes of the Gaussian component.}\\
\tablefoottext{g}{Position angle of the major axis measured from North through East.}
\end{tiny}
}
\end{tiny}
}

\end{document}